\documentclass[useAMS,usenatbib,usegraphicx]{mn2e}

\title{Adiabatic scaling relations of galaxy clusters}
\author[Y.~Ascasibar et al.]
{ Y.~Ascasibar$^1$\thanks{E-mail: yago@aip.de},
  R.~Sevilla$^2$,
  G.~Yepes,$^2$ V.~M\"uller$^1$ and S.~Gottl\"ober$^1$\\
  $^1$Astrophysikalisches Institut Potsdam, An der Sternwarte 16, Potsdam D-14482 (Germany) \\
  $^2$Grupo de Astrof\'\i sica, Universidad Aut\'onoma de Madrid, Madrid E-28049 (Spain)
}
\newcommand{\Referee}[1]{#1}

\newcommand{\LCDM}{$\Lambda$CDM}
\newcommand{\lcdm}{\LCDM\ }

\newcommand{\Msun}{M$_\odot$}
\newcommand{\msun}{\Msun\ }

\newcommand{\be}{\begin{equation}}
\newcommand{\ee}{\end{equation}}
\newcommand{\bea}{\begin{eqnarray}}
\newcommand{\eea}{\end{eqnarray}}

\newcommand{\dd}{{\rm d}}
\newcommand{\cc}{c_\Delta}
\newcommand{\cv}{c_{\rm vir}}
\newcommand{\rs}{r_{\rm s}}
\newcommand{\rhos}{\rho_{\rm s}}

\newcommand{\Rv}{r_{200}}

\newcommand{\Lx}{L_{\rm X}}

\newcommand{\Tx}{T_{\rm X}}

\newcommand{\Ymt}{Y_{\rm MT}}
\newcommand{\Ylt}{Y_{\rm LX}}

\begin{document}

\maketitle

\begin{abstract}
The aim of the present work is to show that, contrary to popular belief, galaxy clusters are \emph{not} expected to be self-similar, even when the only energy sources available are gravity and shock-wave heating.
In particular, we investigate the scaling relations between mass, luminosity and temperature of galaxy groups and clusters in the absence of radiative processes.
Theoretical expectations are derived from a polytropic model of the intracluster medium and compared with the results of high-resolution adiabatic gasdynamical simulations.
It is shown that, in addition to the well-known relation between the mass and concentration of the dark matter halo, the effective polytropic index of the gas also varies systematically with cluster mass, and therefore neither the dark matter nor the gas profiles are exactly self-similar.
It is remarkable, though, that the effects of concentration and polytropic index tend to cancel each other, leading to scaling relations whose logarithmic slopes roughly match the predictions of the most basic self-similar models.
We provide a phenomenological fit to the relation between polytropic index and concentration, as well as a self-consistent scheme to derive the non-linear scaling relations expected for any cosmology and the best-fit normalizations of the M-T, L-T and F-T relations appropriate for a $\Lambda$CDM universe.
The predicted scaling relations \Referee{reproduce observational data reasonably well} for massive clusters, where the effects of cooling and star formation are expected to play a minor role.
\end{abstract}


%--------------------------------------------------------------------------
  \section{Introduction}
  \label{secIntro}
%--------------------------------------------------------------------------

The physics of massive galaxy clusters is relatively simple, at least compared to that of smaller objects.
In the standard cold dark matter (CDM) scenario, their mass is dominated by the dark component, while most baryons are in the form of a hot diffuse plasma in hydrostatic equilibrium with the gravitational potential created by the CDM halo.
The intracluster medium (ICM) gas is shock-heated to approximately the virial temperature of the object, and its thermal bremsstrahlung emission has been detected by X-ray satellites for the last three decades.
In the absence of any other process, it is often stated that galaxy clusters are expected to be self-similar, and their global properties should obey power-law scaling relations.

As long as the shape of a cluster's potential does not depend systematically on its mass, the radial structure of the ICM ought to be scale-free, and the global properties of galaxy clusters, such as halo mass, emission-weighted temperature, or X-ray luminosity, would scale self-similarly \citep{Kaiser86}.
Indeed, numerical simulations that include adiabatic gasdynamics are reported to produce clusters of galaxies that obey such scaling laws \citep[e.g.][]{NFW95,EMN96,BN98,Eke98}.

In real clusters, deviations from self-similarity are expected to arise from merging \citep[e.g.][]{JingSuto00} and additional physics acting on the intracluster gas \citep[see e.g.][and references therein]{TozziNorman01,Babul02,Voit02}.
Radiative cooling and energy injection by supernova and/or active galactic nuclei (AGN) may be particularly relevant for low-mass systems, where they can make a significant contribution to the total energy budget.

Observations seem to corroborate that the self-similar picture is indeed too simplistic, and it fails to predict the observed scalings of cluster mass and luminosity with respect to the ICM gas temperature.

Although some observational studies \citep[e.g.][]{Horner99,NeumannArnaud99,ASF01} are consistent with the self-similar expectation, $M\propto T^{1.5}$, the observed mass-temperature relation has often been found to be steeper \citep[e.g.][]{Sanderson03}, particularly in the group regime \citep[e.g.][]{Nevalainen00,Finoguenov01,Xu01,Arnaud05}.
It has also been noted \citep{EttoriGrandiMolendi02} that the slope of the $M-T$ relation might depend on the limiting overdensity.
Regarding the normalization, the reported value is in most cases $\sim40$ per cent lower than in numerical simulations, although there is a certain degeneracy with the precise value of the slope.

On the other hand, it has been known since the first generation of X-ray satellites \citep[e.g.][]{EdgeStewart91,David93} that the slope of the luminosity-temperature relation, $\alpha\sim3$, is significantly steeper than the self-similar value, $\alpha=2$ \Referee{(although the exact value varies for a limited energy band)}.
It has been shown \citep[e.g.][]{AllenFabian98,Markevitch98,ArnaudEvrard99} that the scatter in the $L-T$ relation is significantly reduced, and the discrepancy is somewhat less severe, when cooling flows are excised or samples with only weak cooling cores are considered.
Actually, it has been recently claimed \citep{OHara_05} that cool core related phenomena, and not merging processes, are the primary contributor to the scatter in all the scaling relations.

Observational data have thus motivated significant efforts attempting to build a physical model of the ICM that breaks self-similarity, either by removing low-entropy gas from the centres of clusters via radiative cooling \citep{Bryan00} or by introducing non-gravitational heating \citep{EvrardHenry91,Kaiser91}.
In both cases, the `excess' entropy produces a flattening of the density profile that brings the X-ray properties of the modelled clusters in agreement with the observed scaling relations \citep[see e.g.][and references therein]{Voit03,Borgani04}.
%\citep[e.g.][and references therein]{Bialek01,Borgani02,Muanwong02,Dave02}.
Nevertheless, the source and precise amount of heating and cooling required are still a matter of debate.
Recent observations \citep[e.g.][]{Ponman03} suggest that the shape of the entropy profile is similar in groups and clusters of galaxies, which rules out the simplest scenarios.

In this paper, we claim that dark matter haloes are not exactly self-similar, and therefore both the cluster's potential and the properties of the ICM gas do indeed depend on the total mass (or temperature) of the object.
In particular, there is no compelling reason to expect that the scaling relations between \Referee{any two physical properties, integrated up to a given overdensity}, should obey a power law, even in the purely gravitational case.

We present a theoretical prediction of these relations based on a polytropic model of the ICM, and compare it with a set of high-resolution adiabatic gasdynamical simulations.
It will be shown that self-similar models implicitly assume that all clusters have the same concentration and polytropic index.
Relaxing these hypotheses yields the scaling relations derived in Section~\ref{secTh}, which we compare with the results of numerical experiments in Section~\ref{secSims}.
Observational implications are discussed in Section~\ref{secObs}, and Section~\ref{secConclus} summarizes our main conclusions.

%--------------------------------------------------------------------------
  \section{Theoretical model}
  \label{secTh}
%--------------------------------------------------------------------------

As shown by \citet{Ascasibar03}, relaxed clusters and minor mergers found in adiabatic gasdynamical simulations can be considered to be in approximate thermally-supported hydrostatic equilibrium up to $\sim0.8\Rv$.
Furthermore, the ICM gas is fairly well described by a polytropic equation of state with an effective polytropic index $\gamma\simeq1.18$.
Using the phenomenological formula proposed by \citet[hereafter NFW]{NFW97} to model the density profile of the dark matter halo,
\be
\rho(r)=\frac{\rhos}{(r/\rs)(1+r/\rs)^2},
\label{eqNFW}
\ee
the gas temperature is given by
\be
T(r) = T_0\ \frac{\ln(1+r/\rs)}{r/\rs},
\label{eqTemp}
\ee
where the central temperature,
\be
kT_0= 4\pi G\mu m_{\rm p} \frac{\gamma-1}{\gamma}\rhos \rs^2,
\label{eqT0}
\ee
is set by the boundary condition that the ICM density and temperature vanish at infinity.
The gas density profile can be computed from the polytropic relation
\be
\rho_{\rm g}(r) = \rho_0(\gamma)
    \left[ \frac{\ln(1+r/\rs)}{r/\rs} \right]^{\frac{1}{\gamma-1}},
\label{eqGas}
\ee
where the central gas density, $\rho_0(\gamma)$, can be constrained by normalizing the baryon fraction to match the cosmic value at $x_{\rm b}=r_{\rm b}/\rs\sim3$,
\be
\rho_0(\gamma)=\frac{\Omega_{\rm b}}{\Omega_{\rm dm}}\,\rhos
    \left[ \frac{\ln(1+x_{\rm b})}{x_{\rm b}} \right]^{-\frac{1}{\gamma-1}}
    \left[ x_{\rm b}(1+x_{\rm b})^2 \right]^{-1}.
\ee
This analytic prescription is simpler than the one proposed in \citet{Ascasibar03}, and it provides better results for extreme values of the polytropic index $\gamma$.
The choice $x_{\rm b}\sim3$ is somewhat arbitrary, and in principle one could leave the normalization of the gas density as a free parameter of the model.

We consider, though, that it is desirable to reduce the number of free parameters as much as possible.
In fact, the very existence of relatively tight scaling relations suggests that real galaxy clusters can indeed be described by only one free parameter, which could be taken to be the mass of the halo.
Once $x_{\rm b}$ is set, our model still has three parameters, two of them related to the dark matter halo (the characteristic density and radius, $\rhos$ and $\rs$) and one related to the intracluster gas (the effective polytropic index $\gamma$).
The first two are known to be correlated \citep{NFW97,Bullock01,Eke01}, and we propose a phenomenological relation between polytropic index and concentration in Section~\ref{secSims} below.

It is important to note, though, that the fact that clusters could be described by a one-parameter family of functions would give rise to universal scaling relations both for their radial profiles \Referee{as well as for their global (integrated or averaged) physical properties}.
However, it does \emph{not} imply that such relations ought to be self-similar in any sense.
The precise functional form of the different scalings would be specified by the two independent relations between $\rhos$, $\rs$ and $\gamma$.

In the present work, we are interested in the scaling relations between several quantities, integrated up to the radius $r_\Delta$ encompassing an overdensity $\Delta$ with respect to the critical density, i.e.
\be
M_\Delta \equiv M(r_\Delta) = \Delta\frac{4\pi}{3}\rho_{\rm c}\,r_\Delta^3,
\ee
where $\rho_{\rm c}\equiv\frac{3H_0^2}{8\pi G}=2.8\times10^{11}\ h^2$ \msun Mpc$^{-3}$ is the critical density and $H_0\equiv 100\ h$ km s$^{-1}$ Mpc$^{-1}$ is the Hubble constant.
Defining $\cc \equiv r_\Delta / \rs$ and $g(x)\equiv\left[\ln(1+x)-\frac{x}{1+x} \right]^{-1}$, we obtain
\be
M_\Delta \simeq M_{\rm dm}^\Delta = 4\pi\rhos\rs^3\ g^{-1}(\cc)
\label{eqMs}
\ee
and
\be
\rhos = \frac{\Delta H_0^2}{8\pi G}\,\cc^3\ g(\cc).
\label{eqRhos}
\ee

Let us define the function
\be
y(\eta,c)\equiv
  \int_0^{c}\left[ \frac{\ln(1+x)}{x} \right] ^{\eta} x^2\ \dd x,
\ee
which must be integrated numerically.
In terms of this function, we can express in a very compact form both the cumulative gas mass,
\be
M_{\rm g}^\Delta = 4\pi\rho_0\rs^3\ y(\frac{1}{\gamma-1},\cc),
\label{eqMgas}
\ee
and mass-weighted temperature,
\be
T_\Delta=T_0\,
\frac{ y(\frac{\gamma}{\gamma-1},\cc)}
     { y(\frac{   1  }{\gamma-1},\cc)}.
\label{eqT}
\ee

Assuming that thermal bremsstrahlung is the dominant cooling mechanism, the X-ray power radiated by the ICM gas per unit volume may be estimated as
\be
P=\frac{64\pi}{3}\left(\frac{\pi}{6}\right)^{\!\frac{1}{2}}
   \left(\frac{e^2}{4\pi\epsilon_0}\right)^{\!\!3}
   \left[\frac{kT}{(m_{\rm e}c^2)^3}\right]^{\!\frac{1}{2}}
   \bar{g} n_{\rm e}\sum_i{Z_i^2n_i},
\ee
where $e$, $m_{\rm e}$ and $n_{\rm e}$ are the electron charge, mass and number density, respectively.
$\epsilon_0$ is the permittivity of free space, $k$ is Boltzmann's constant, $c$ is the speed of light and $\bar{g}$ is the average Gaunt factor, which we take to be unity.
The sum takes into account the atomic number $Z_i$ and number density $n_i$ of each ion species $i$.
For a fully-ionized plasma of primordial composition ($\sim75$ per cent of the mass in hydrogen and 25 per cent in helium),
\be
P \simeq 2\times10^{17}
   \left(\!\!\frac{\rho}{\rm M_\odot\ Mpc^{-3}}\!\right)^{\!2}
   \left(\frac{kT}{\rm keV}\right)^{\!\frac{1}{2}}
{\rm \ erg\ s^{-1}\,Mpc^{-3}.}
\ee
Integrating up to $r_\Delta$, the bolometric X-ray luminosity would be
\be
\Lx^\Delta = \Lambda_{\rm X}\ \rho_0^2\ (kT_0)^{\!\frac{1}{2}}
   \,4\pi\rs^3 \ y(\frac{2}{\gamma-1}\!+\!\frac{1}{2},\cc)
\label{eqLx}
\ee
with $\Lambda_{\rm X}\simeq2\times10^{17}$ erg s$^{-1}$ Mpc$^{3}$ \Msun$^{-2}$ keV$^{-\frac{1}{2}}$, while the emission-weighted temperature is given by
\be
\Tx^\Delta=T_0\,
\frac{ y(\frac{2}{\gamma-1}+\frac{3}{2},\cc)}
     { y(\frac{2}{\gamma-1}+\frac{1}{2},\cc)}.
\label{eqTx}
\ee

Combining equations (\ref{eqT0}), (\ref{eqMs}), (\ref{eqRhos}) and (\ref{eqTx}), simple algebra yields the mass-temperature relation
\be
M_\Delta =
    \frac{\sqrt{2}}{GH_0}
    \Delta^{-1/2}
    \left[\frac{k\Tx^\Delta}{\mu m_p}\right]^{3/2}
    \Ymt(\gamma,\cc),
\label{eqMT}
\ee
where
\be
\Ymt(\gamma,\cc) \equiv
\left[
  \frac{\gamma-1}{\gamma}\,\cc\ g(\cc)\,
  \frac{ y(\frac{2}{\gamma-1}+\frac{3}{2},\cc)}
       { y(\frac{2}{\gamma-1}+\frac{1}{2},\cc)}\,
\right]^{-\frac{3}{2}}.
\label{eqYmt}
\ee

Analogously, expressions (\ref{eqMs}) and (\ref{eqMgas}) tell us that the cumulative baryon fraction does not depend explicitly on the object mass or temperature,
\be
F_\Delta \equiv \frac{M_{\rm g}^\Delta}{M_\Delta} \frac{\Omega_{\rm dm}}{\Omega_{\rm b}}=
    \frac{\rho_0(\gamma)}{\rhos}\frac{\Omega_{\rm dm}}{\Omega_{\rm b}}\,
    g(\cc)\ y(\frac{1}{\gamma-1},\cc).
\label{eqFb}
\ee

Finally, the luminosity-temperature relation can be expressed in the form
\be
\Lx^\Delta=
    \frac{\Lambda_{\rm X}H_0}{(\mu m_{\rm p})^{\frac{3}{2}}\,2^{\frac{5}{2}}\,\pi\,G^2}
    \left(k\Tx^\Delta\right)^2 \Delta^{\frac{1}{2}}\ \Ylt(\gamma,\cc)
\label{eqLT}
\ee
with
\be
\Ylt(\gamma,\cc)\!\equiv\!\!\!
  \left[\!\frac{\rho_0\!(\gamma)}{\rhos}\!\right]^2\!
  \left[\!\frac{\gamma\,\cc}{\gamma-1}\!\right]^{\!-\frac{3}{2}}
  \!\!g^{\frac{1}{2}}\!(\cc)\,
  \frac{ y^3(\frac{2}{\gamma-1}+\frac{1}{2},\cc)}
       { y^2(\frac{2}{\gamma-1}+\frac{3}{2},\cc)},
\label{eqYlt}
\ee
according to (\ref{eqT0}), (\ref{eqRhos}), (\ref{eqLx}) and (\ref{eqTx}).

%--------------------------------------------------------------------------
  \section{Simulations}
  \label{secSims}
%--------------------------------------------------------------------------

For constant values of the polytropic index and concentration, equations (\ref{eqMT}), (\ref{eqFb}) and (\ref{eqLT}) become the well-known self-similar scalings, with logarithmic slopes $3/2$, $0$ and $2$, respectively.
The precise `universal' values of $\gamma$ and $\cc$ would simply set the normalization.

However, both quantities might well depend systematically on the mass of the object.
We address such dependence in the present section, where we also compute the expected scaling relations and compare them to our numerical data.

%__________________________________
\begin{table*}
  \caption{
Description of our cluster sample.
Number of gas particles within $\Rv$, \Referee{physical properties} at overdensities $\Delta=2500$, 500 and 200, best-fitting characteristic density (in units of the critical density), radius (in $h^{-1}$ kpc) and effective polytropic index.
Masses are expressed in $10^{13}$ \Msun, temperatures in keV and X-ray luminosities in $10^{44}\ h$ erg s$^{-1}$.
The baryon fraction is given in units of the cosmic value, $\Omega_{\rm b}/\Omega_{\rm m}$.
}
\label{tabSims}
\centering
\begin{tabular}{rr rrrr rrrr rrrr rrr}
\hline
 $N_{\rm gas}$ &
    $M_{2500}$ & $F_{2500}$ & $\Tx^{2500}$ & $\Lx^{2500}$
 &  $M_{ 500}$ & $F_{ 500}$ & $\Tx^{ 500}$ & $\Lx^{ 500}$
 &  $M_{ 200}$ & $F_{ 200}$ & $\Tx^{ 200}$ & $\Lx^{ 200}$
 &    $\rhos/\rho_{\rm c}$ & $\rs$~ & $\gamma$~~ \\
\hline
 1511675 & 50.33 & 0.60 & 10.60 & 21.44  &   118.35 & 0.85 & 10.11 & 29.67  &   177.53 & 0.90 &  9.88 & 32.28  &    5970 &   434 &   1.170 \\
   94804 &  3.84 & 0.78 &  2.17 &  1.65  &     8.42 & 0.83 &  2.14 &  1.81  &    12.34 & 0.81 &  2.17 &  1.85  &   12698 &   136 &   1.176 \\
   71366 &  2.58 & 0.66 &  1.75 &  0.43  &     6.41 & 0.83 &  1.62 &  0.60  &     8.85 & 0.85 &  1.61 &  0.62  &    5792 &   177 &   1.170 \\
 1191370 & 34.27 & 0.89 & 10.62 & 28.44  &    92.44 & 0.93 & 10.51 & 36.29  &   136.74 & 0.92 & 10.38 & 37.47  &    4255 &   509 &   1.154 \\
 1089303 & 16.94 & 0.53 & 11.66 &  8.24  &    82.84 & 0.85 & 10.64 & 34.74  &   120.72 & 0.95 & 10.42 & 37.03  &    1139 &   928 &   1.158 \\
  970021 & 30.13 & 0.75 &  8.58 & 15.25  &    75.78 & 0.94 &  8.14 & 20.64  &   113.58 & 0.90 &  8.04 & 21.65  &    7505 &   346 &   1.167 \\
  836640 &  2.87 & 0.52 &  3.31 &  0.35  &    30.11 & 0.74 &  6.14 &  3.56  &    99.79 & 0.88 &  8.05 & 21.23  &     636 &   855 &   1.158 \\ 
  740358 & 16.97 & 0.59 &  7.18 &  4.04  &    52.51 & 0.87 &  5.74 &  7.98  &    90.08 & 0.87 &  5.60 &  8.80  &   11094 &   243 &   1.194 \\
 1325010 & 41.67 & 0.70 & 11.82 & 19.68  &   105.05 & 0.88 & 10.99 & 27.40  &   153.94 & 0.91 & 10.82 & 28.53  &    5343 &   465 &   1.167 \\
  124201 &  5.50 & 0.42 &  3.84 &  0.49  &    11.40 & 0.70 &  3.29 &  0.76  &    15.95 & 0.82 &  3.17 &  0.82  &   11098 &   161 &   1.221 \\
  276238 & 11.21 & 0.93 &  4.46 & 11.59  &    22.60 & 0.95 &  4.40 & 12.30  &    30.70 & 0.95 &  4.39 & 12.38  &   18060 &   163 &   1.164 \\
  670893 & 22.94 & 0.81 &  7.60 & 12.96  &    51.66 & 0.94 &  7.38 & 15.77  &    73.35 & 0.96 &  7.31 & 16.18  &    5002 &   394 &   1.159 \\
   27706 &  1.39 & 0.43 &  1.58 &  0.08  &     2.94 & 0.63 &  1.43 &  0.11  &     4.19 & 0.70 &  1.43 &  0.12  &    6786 &   136 &   1.217 \\
  304438 &  3.66 & 0.88 &  1.90 &  1.57  &     8.07 & 0.87 &  1.92 &  1.80  &     9.59 & 0.88 &  1.91 &  1.81  &    8634 &   161 &   1.164 \\
  134337 &  1.44 & 0.70 &  1.23 &  0.23  &     3.08 & 0.83 &  1.17 &  0.27  &     4.16 & 0.89 &  1.15 &  0.28  &    9156 &   114 &   1.179 \\
  123954 &  1.62 & 0.74 &  1.36 &  0.43  &     2.95 & 0.87 &  1.32 &  0.47  &     3.66 & 0.93 &  1.31 &  0.47  &   30553 &    65 &   1.202 \\
   86035 &  1.02 & 0.67 &  1.03 &  0.17  &     2.16 & 0.78 &  0.99 &  0.19  &     2.80 & 0.85 &  0.99 &  0.20  &   14820 &    81 &   1.198 \\
  120196 &  0.48 & 0.69 &  0.65 &  0.05  &     2.27 & 0.80 &  0.68 &  0.11  &     4.30 & 0.83 &  0.67 &  0.14  &    1085 &   285 &   1.152 \\
  119735 &  0.46 & 0.47 &  0.70 &  0.02  &     2.20 & 0.79 &  0.69 &  0.10  &     4.29 & 0.83 &  0.67 &  0.14  &     921 &   311 &   1.152 \\
   81150 &  1.02 & 0.85 &  1.00 &  0.46  &     1.97 & 0.90 &  0.98 &  0.48  &     2.57 & 0.94 &  0.98 &  0.48  &   26165 &    62 &   1.190 \\
  134696 &  1.88 & 0.76 &  1.56 &  0.53  &     3.48 & 0.86 &  1.51 &  0.58  &     4.33 & 0.92 &  1.50 &  0.58  &   30406 &    69 &   1.194 \\
   73495 &  0.51 & 0.41 &  0.64 &  0.02  &     1.17 & 0.67 &  0.64 &  0.03  &     2.80 & 0.79 &  0.61 &  0.05  &    1471 &   225 &   1.166 \\
   17632 &  0.25 & 0.43 &  0.44 &  0.01  &     0.49 & 0.61 &  0.43 &  0.01  &     0.66 & 0.81 &  0.42 &  0.01  &    2653 &   124 &   1.218 \\
  241690 &  3.27 & 0.75 &  1.95 &  0.91  &     6.14 & 0.86 &  1.88 &  1.02  &     8.20 & 0.88 &  1.87 &  1.03  &   17239 &   107 &   1.175 \\
   98380 &  0.93 & 0.52 &  0.89 &  0.06  &     2.14 & 0.76 &  0.81 &  0.09  &     3.62 & 0.82 &  0.78 &  0.10  &    4260 &   146 &   1.172 \\
 1400058 &  5.31 & 0.65 &  4.01 &  1.17  &    28.13 & 0.80 &  3.46 &  2.83  &    45.73 & 0.85 &  3.31 &  3.18  &    2208 &   422 &   1.161 \\
  111187 &  1.33 & 0.73 &  1.12 &  0.24  &     2.62 & 0.90 &  1.06 &  0.28  &     3.74 & 0.88 &  1.04 &  0.29  &   14668 &    88 &   1.176 \\
   76770 &  0.92 & 0.63 &  0.89 &  0.15  &     1.71 & 0.85 &  0.83 &  0.17  &     2.68 & 0.85 &  0.81 &  0.18  &    9841 &    92 &   1.167 \\
  185292 &  1.99 & 0.74 &  1.21 &  0.44  &     5.16 & 0.87 &  1.20 &  0.72  &     6.36 & 0.86 &  1.20 &  0.73  &    4798 &   174 &   1.158 \\
  179971 &  1.31 & 0.70 &  1.19 &  0.26  &     4.82 & 0.83 &  1.21 &  0.71  &     6.13 & 0.87 &  1.20 &  0.73  &    2866 &   221 &   1.162 \\
  163595 &  0.86 & 0.61 &  0.93 &  0.08  &     3.39 & 0.79 &  0.86 &  0.18  &     5.66 & 0.86 &  0.82 &  0.26  &    1689 &   260 &   1.155 \\
  162520 &  0.62 & 0.69 &  0.74 &  0.07  &     2.89 & 0.80 &  0.79 &  0.16  &     5.62 & 0.86 &  0.82 &  0.26  &    1052 &   321 &   1.150 \\
  934064 & 10.82 & 0.76 &  4.41 &  4.61  &    22.15 & 0.87 &  4.23 &  5.43  &    29.11 & 0.89 &  4.20 &  5.51  &   10437 &   205 &   1.169 \\
  897423 &  5.53 & 0.76 &  4.58 &  2.41  &    20.97 & 0.86 &  4.24 &  5.38  &    28.28 & 0.88 &  4.21 &  5.49  &    2260 &   429 &   1.155 \\
  132470 &  1.64 & 0.71 &  1.32 &  0.40  &     2.97 & 0.86 &  1.28 &  0.44  &     3.85 & 0.95 &  1.27 &  0.44  &   21423 &    77 &   1.202 \\
  416239 &  5.28 & 0.58 &  2.85 &  0.91  &    11.20 & 0.77 &  2.66 &  1.15  &    14.46 & 0.85 &  2.62 &  1.18  &    9667 &   168 &   1.183 \\
  501065 &  5.28 & 0.83 &  2.82 &  3.03  &    11.67 & 0.88 &  2.75 &  3.32  &    15.75 & 0.88 &  2.73 &  3.36  &   18629 &   127 &   1.171 \\
 4490660 &  5.95 & 0.88 &  3.13 &  4.95  &    12.02 & 0.92 &  3.07 &  5.18  &    17.18 & 0.90 &  3.05 &  5.25  &   27337 &   109 &   1.168 \\
  436426 &  5.78 & 0.80 &  2.90 &  2.54  &    11.08 & 0.88 &  2.83 &  2.75  &    14.42 & 0.90 &  2.81 &  2.77  &   17751 &   129 &   1.173 \\
   59817 &  0.84 & 0.67 &  0.82 &  0.13  &     1.94 & 0.67 &  0.80 &  0.14  &     2.70 & 0.66 &  0.80 &  0.14  &   12304 &    82 &   1.181 \\
  274209 &  2.52 & 0.72 &  1.80 &  0.50  &     6.39 & 0.82 &  1.72 &  0.64  &     9.18 & 0.83 &  1.69 &  0.66  &    6565 &   167 &   1.171 \\
  157916 &  2.17 & 0.82 &  1.57 &  0.86  &     3.92 & 0.88 &  1.55 &  0.91  &     4.76 & 0.92 &  1.54 &  0.91  &   26205 &    77 &   1.185 \\
\hline
\end{tabular}
\end{table*}
%__________________________________

%___________________________________________________
\subsection{Numerical experiments}

Our cluster sample consists of 42 objects formed in a flat \lcdm universe ($\Omega_{\rm m}=0.3$; $\Omega_{\rm b}=0.04$; $\Omega_\Lambda=0.7$; $ h=0.7$; $\sigma_8=0.9$).
28 of them have been extracted from a $80~h^{-1}$~Mpc cubic box simulated with a version of the parallel Tree-SPH code {\sc Gadget} \citep{gadget01} that implements the entropy-conserving scheme proposed by \citet{Gadget02}.
For a thorough description of these experiments, the reader is referred to \citet{tesis}.
In order to extend our numerical sample of clusters to a wider temperature (mass) range, we have also simulated a $500~h^{-1}$~Mpc box with the code {\sc Gadget2} \citep{gadget2}, from which we have considered 14 objects.
Details about these simulations can be found in \citet{Yepes_04}.

In each case, high resolution has been achieved by means of the multiple-mass technique \citep[see][]{Klypin01}.
An unconstrained random realization of the \lcdm power spectrum was generated with $1024^3$ and $2048^3$ particles for the 80 and $500~h^{-1}$~Mpc boxes, respectively.
Haloes were selected at $z=0$ from a low-resolution experiment evolved with $128^3$ dark matter particles, and then re-simulated with three and five levels of mass refinement (so that the final mass resolution of both subsamples is similar).
The gravitational softening length was set to $\epsilon=2-5\ h^{-1}$~kpc, depending on number of dark matter particles within the virial radius of the object \citep[following][]{Power03}.
Gas particles have only been added in the highest refinement level.

Basic information about the objects in our numerical cluster sample is summarized in Table~\ref{tabSims}.
They span two orders of magnitude in mass ($\sim10^{13}-10^{15}$~\Msun) and cover a temperature range between 0.5 and 11 keV.
The total number of gas particles within the virial radius is always $N_{\rm gas}\ge2\times10^4$, the number of dark matter particles being slightly higher (inversely proportional to the cumulative baryon fraction, $F$).

For each object, the centre of mass was found by an iterative procedure.
Starting with an initial guess, we compute the centre of mass within a sphere of $500\ h^{-1}$ kpc.
The sphere is moved to the new centre until convergence is reached.
The radius of the sphere is then decreased by 10 per cent, and the process continues until the sphere contains 200 dark matter particles.
The radii $r_{2500}$, $r_{500}$ and $r_{200}$ are obtained
from the overdensity profile around the final centre of mass.
The total mass, baryon fraction, X-ray luminosity and emission-weighted temperature quoted in Table~\ref{tabSims} have been computed within those radii.

%___________________________________________________

\subsection{Concentration and polytropic index}

The values of the parameters $\cc$ and $\gamma$ have been computed by means of a \Referee{global fit} to the gas density, dark matter density, total mass and effective polytropic index (i.e. gas temperature versus gas density) profiles, averaged over logarithmically-spaced spherical shells between $0.1\Rv$ and $\Rv$.
\Referee{
More precisely, we minimize the quantity
\be
\chi^2=\chi^2_{\rho}+\chi^2_{M}+\chi^2_{T}+\chi^2_{\gamma},
\ee
where
\be
\chi^2_{x}=\sum_{b=1}^{n_{\rm bins}}\frac{\log^2[\,x(b)/x_{\rm model}(b)\,]}{1/{n(b)}}
\ee
and $x$ denotes the number of particles within each bin, $n(b)$, the total enclosed mass, $M(b)$, the average temperature within the bin, $T(b)=\sum_{i=1}^{n(b)}T_i/n(b)$, and the quantity $T(b)n^{1-\gamma}(b)$.
}

A grid of analytical profiles is generated for the intervals $1.1<\gamma<1.25$ and $10<\rs/(h^{-1}{\rm kpc})<1000$, in uniform steps $\Delta\gamma=0.001$ and $\Delta\rs=1\ h^{-1}$ kpc.
Since all the measured quantities are proportional to the characteristic density $\rhos$, its best-fitting value has been trivially found from the average of the logarithmic residuals.

%__________________________________
\begin{figure}
  \centering \includegraphics[width=8cm]{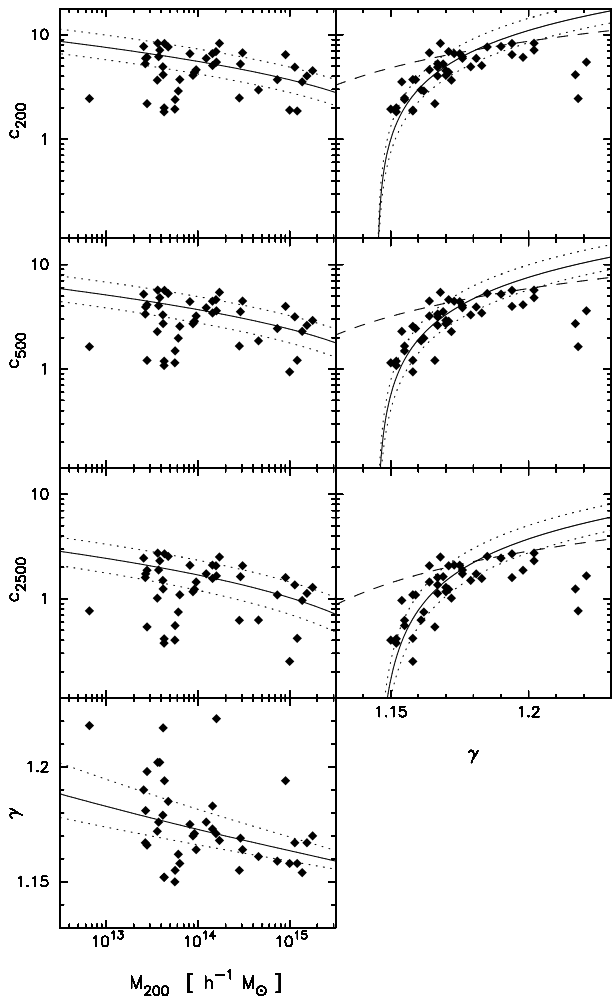}
  \caption{
Concentration at different overdensities and effective polytropic index of each object.
Solid lines represent our theoretical model, based on the mass-concentration relation of \citet{Bullock01} and our proposed fit to the dependence of the effective polytropic index on concentration, equation~(\ref{eqCG}).
Dotted lines show the one-sigma scatter expected from $\Delta\cv/\cv\simeq0.3$ \citep{Colin04}.
\Referee{The linear fit to the $\gamma-c$ relation proposed by \citet{KS01} is shown by the dashed lines on the right panels}.
}
  \label{figCG}
\end{figure}
%__________________________________

Results of the minimization procedure are quoted in Table~\ref{tabSims}, and best-fitting concentrations and effective polytropic indices are plotted in Figure~\ref{figCG}.
Solid lines depict the toy model for the mass-concentration relation proposed by \citet{Bullock01}\footnote{Nearly identical results are obtained when the prescription given in \citet{Eke01} is used.}, with $F=0.001$ and $K=3$.
We transform the values of $\cv\simeq c_{100}$ to the other overdensities according to the NFW profile.
Dotted lines show the one-sigma scatter $\Delta\cv/\cv\simeq0.3$ reported by \citet{Colin04} for relaxed systems.
Finally, we find that the phenomenological relation
\be
\gamma=a+b\,c_{200}
\label{eqCG}
\ee
\Referee{
with $a=1.145\pm0.007$ and $b=0.005\pm0.002$ fits reasonably well our results for the polytropic index (although there exists a certain degeneracy between the best-fitting values of both parameters).
Dotted lines show the one-sigma scatter in $\gamma(c)$ expected from $\Delta\cv$.

A correlation between $\gamma$ and $c$ is expected if both quantities (and therefore the radial structure of the ICM) vary smoothly with the mass of the object.
In fact, an approximately linear dependence has already been advocated by \citet{KS01} in order to enforce constant baryon fraction at large radii.
Their fit (dashed lines in Figure~\ref{figCG}) is however significantly steeper than equation (\ref{eqCG}).
Although it works somewhat better for large $c$, it does not seem to adequately describe the least concentrated systems, suggesting that, most probably, the precise functional form of $\gamma(c)$ is not as simple as a straight line.
We therefore advise against extrapolating our fit towards values of the concentration parameter outside the range covered by the present work.
Moreover, additional physics is expected to play an important role in less massive (more concentrated) systems, and therefore the polytropic approximation itself will no longer to be valid, since it fails to describe the presence of a central cool core.
}

Some of our objects deviate appreciably from both the mass-concentration relation and the $\gamma-c$ relation given by expression (\ref{eqCG}).
These tend to be merging systems, which have formed (or \emph{are} forming) more recently than relaxed objects.
In the spherical collapse picture, that means they have collapsed around density peaks on larger scales, and therefore the resulting density profiles are less concentrated than relaxed haloes of the same mass \citep{Ascasibar04}.
As noted by \citet{Gottloeber01}, merging is more common on the scale of galaxy groups.
The effective polytropic index seems to be systematically higher in these objects, although it is important to bear in mind that they are not particularly well described by a polytropic equation of state.
\Referee{
Actually, the gas distribution shows obvious asymmetries, as well as an offset between the gas and dark matter peak which results in an artificially flat gas density profile in the central regions.
Such flattening is responsible for both the abnormally low baryon fraction measured at $\Delta=2500$ and the unusually high value of $\gamma$ obtained by our fitting routine.
}

%___________________________________________________

\subsection{Scaling relations}

%__________________________________
\begin{figure*}
  \centering \includegraphics[width=15cm]{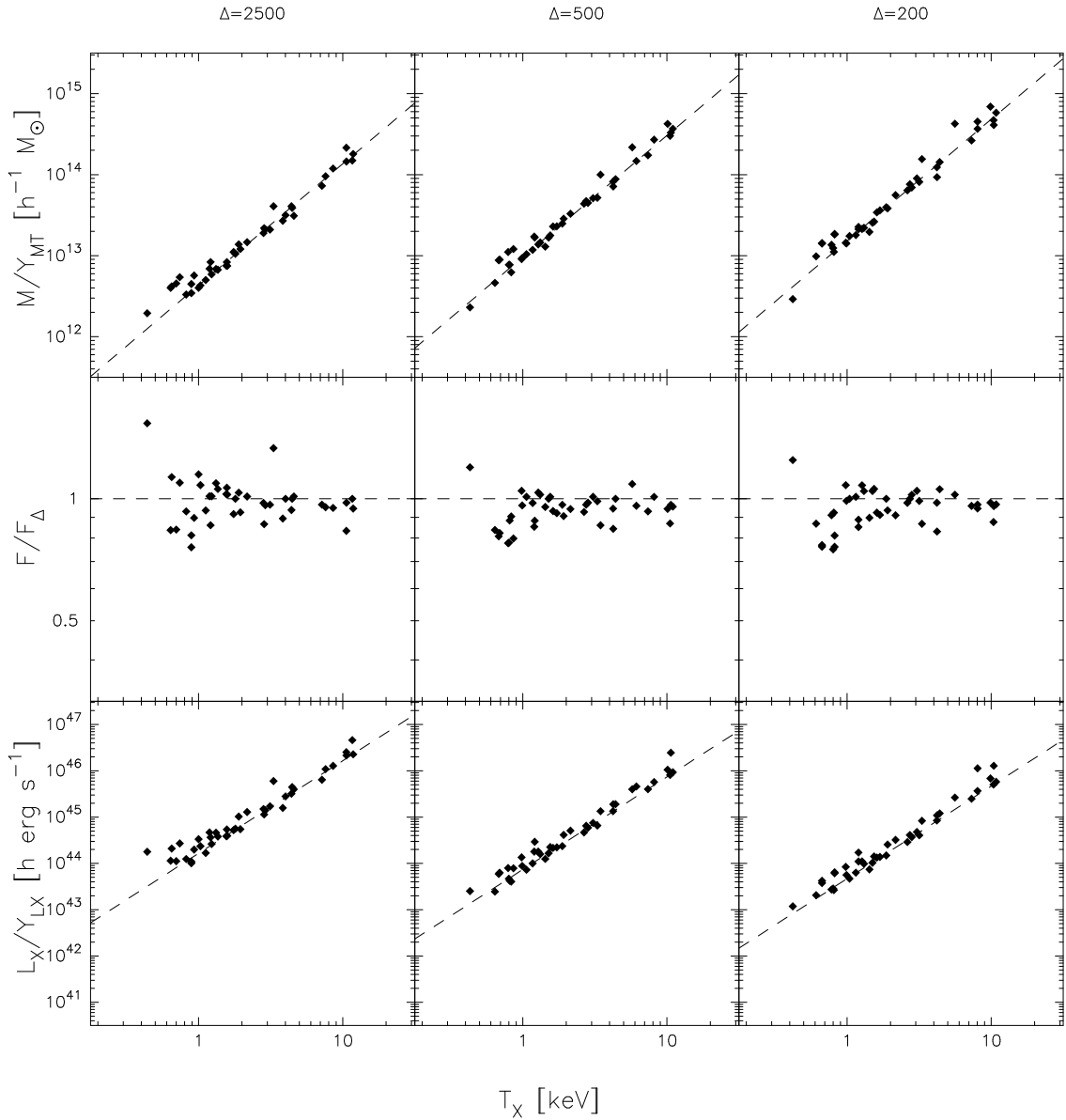}
  \caption{
$M-\Tx$, $F-\Tx$ and $\Lx-\Tx$ scaling relations of our cluster sample at different overdensities, corrected by the factors $\Ymt$, $F_\Delta$ and $\Ylt$, respectively.
Dashed lines represent the theoretical predictions given by expressions (\ref{eqMT}), (\ref{eqFb}) and (\ref{eqLT}).
}
  \label{figCorr}
\end{figure*}
%__________________________________

The scaling relations of total mass, baryon fraction and X-ray luminosity with respect to the emission-weighted temperature are represented in Figure~\ref{figCorr}, divided by the appropriate values of the structure factors $\Ymt$, $F_\Delta$ and $\Ylt$ corresponding to each object.
These are computed by substituting the best-fitting values of $\gamma$ and $\cc$ into expressions (\ref{eqYmt}), (\ref{eqFb}) and (\ref{eqYlt}).

As shown by \citet{Ascasibar03}, polytropic models provide a fairly accurate description of the radial structure of galaxy groups and clusters.
It is therefore not surprising that they are able to match the scaling relations as well.
When the factors $\Ymt$, $F_\Delta$ and $\Ylt$ are taken into account, both the normalization and the logarithmic slope (equal to the self-similar models) of the scaling relations are correctly predicted by equations (\ref{eqMT}), (\ref{eqFb}) and (\ref{eqLT}).
The scatter around the theoretical expectation is quite low, and only merging systems deviate appreciably from the predicted relation.
In these objects, the dark matter potential may differ considerably from the NFW form, and the assumptions of hydrostatic equilibrium and a polytropic equation of state provide rather poor approximations \citep{Ascasibar03}.

%__________________________________
\begin{figure*}
  \centering \includegraphics[width=15cm]{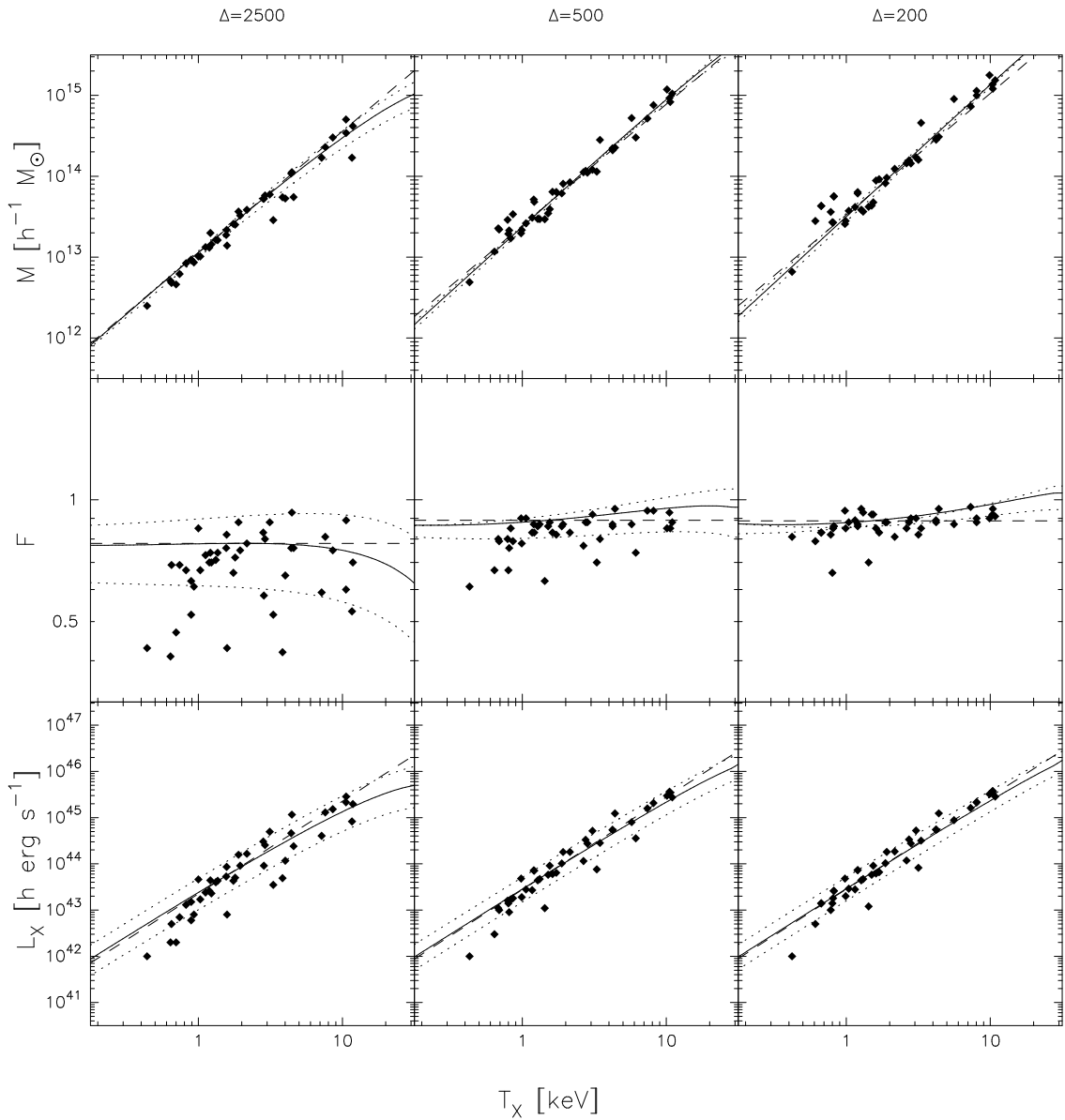}
  \caption{
    Numerical scaling relations (points), compared to our theoretical prediction (solid lines) based on a variable $c(M)$ and $\gamma(c)$, see text.
    Dotted lines show the scatter arising from $\Delta\cv/\cv\simeq0.3$, and dashed lines indicate the scaling relations expected for a fixed concentration $\cv=8$ and effective polytropic index $\gamma=1.176$.
  }
  \label{figUncorr}
\end{figure*}
%__________________________________

Uncorrected scaling relations are plotted in Figure~\ref{figUncorr}.
Solid lines show our theoretical prediction, using the mass-concentration relation from \citet{Bullock01} and our fit (\ref{eqCG}) to estimate $c_\Delta$ and $\gamma$ as a function of the cluster mass.
It turns out that the dependencies on concentration and polytropic index seem to cancel each other so that the resulting scaling relations look as if the clusters were actually self-similar.
Indeed, all the relations considered in the present study can be accurately fit by an `average' concentration $\cv=8$ (which implies $c_{200}\simeq6$, $c_{500}\simeq4$, $c_{2500}\simeq1.8$ and $\gamma\simeq1.176$).
The corresponding scaling relations,
\bea
M_\Delta &=& M_0^\Delta\Delta^{-1/2}\,\left(\frac{\Tx}{1~{\rm keV}}\right)^{3/2}\\
F_\Delta &=& F_0^\Delta \\
\Lx^\Delta &=& L_0^\Delta\left(\frac{\Tx}{1~{\rm keV}}\right)^2,
\eea
 have been plotted as dashed lines, and their normalizations are given in Table~\ref{tabFiducial}.

%__________________________________
\begin{table}
\caption{
Normalization of the approximate scaling relations obtained for $\cv=8$ and $\gamma=1.176$.
$M_0^\Delta$ is expressed in $10^{14}~h^{-1}$~\Msun, $L_0^\Delta$ in $10^{43}~h$~erg~s$^{-1}$ and $F_0^\Delta$ in units of the cosmic baryon fraction.
}
 \centering
 \label{tabFiducial}
 \begin{tabular}{cccc}
 \hline
 $\Delta$ & $M_0^\Delta$ & $F_0^\Delta$ & $L_0^\Delta$\\
    \hline
  2500 & 5.65 & 0.771 & 2.18 \\
  500  & 5.47 & 0.885 & 2.69 \\
  200  & 4.71 & 0.885 & 2.74 \\
\hline
\end{tabular}
\end{table}
%__________________________________

%__________________________________
\begin{figure*}
  \centering \includegraphics[width=15cm]{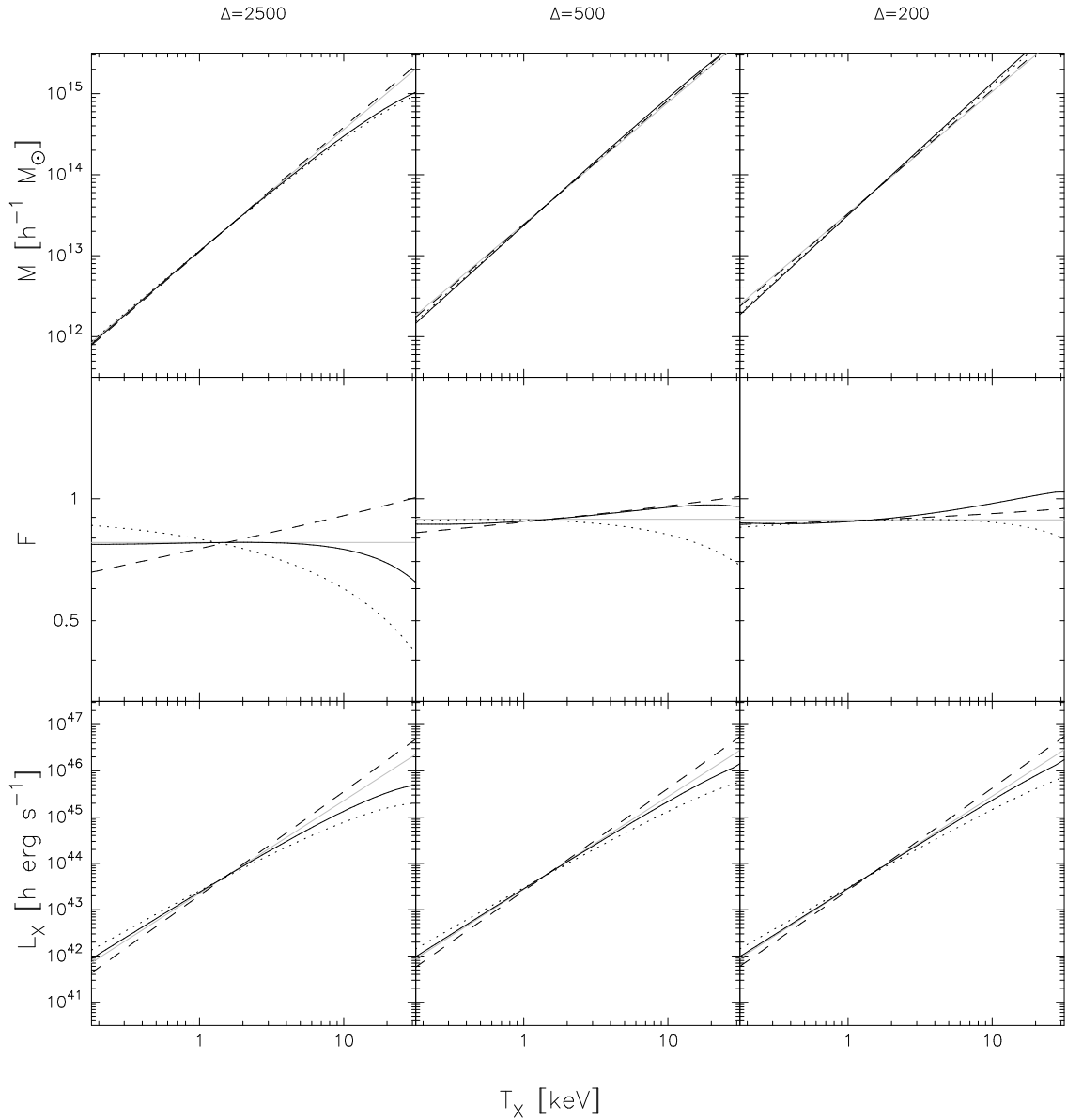}
  \caption{
\Referee{
Effects of concentration (dotted) and polytropic index (dashed) on initially self-similar scaling relations where $\cv=8$ and $\gamma=1.176$ for all objects (grey solid).
Black solid lines show our prediction, taking into account both effects simultaneously.
}
  }
  \label{figCancel}
\end{figure*}
%__________________________________

\Referee{
It is to some extent remarkable that concentration and polytropic index conspire to produce scaling relations that match so closely the self-similar slope.
Such effect is illustrated more clearly in Figure~\ref{figCancel}, where the contributions of $\gamma$ and $c$ to the scaling relations are plotted separately.

The value of the concentration sets the ratio between the radii $r_\Delta$ and the characteristic radius $\rs$.
Since massive objects are substantially less concentrated than smaller systems, their $r_\Delta$ are much closer to the centre in terms of $\rs$, and thus the mass within a given overdensity (which is an increasing function of $r/\rs$) will be smaller than indicated by the `average' scaling relation based on a higher value of $\cv$.
Conversely, $M_\Delta$ in the least massive objects would be biased high with respect to the self-similar relation.
The effect is particularly noticeable for $\Delta=2500$, where $r_\Delta\ll \rs$ and the enclosed mass is a rapidly increasing function of $r/\rs$; for $r>\rs$, it increases only logarithmically, and differences in $r_\Delta/\rs$ are much less important.

The emission-weighted temperature decreases with $r/\rs$, and therefore it is expected to be biased high (low) for large (small) systems.
However, the effect is arguably small, since the average is biased towards the central part, where the gas temperature is roughly constant.
As shown by the dotted lines on the top panels of Figure~\ref{figCancel}, the net result is a shallower mass-temperature relation at $\Delta=2500$, while no significant change can be appreciated at lower overdensities.

The baryon fraction and the X-ray luminosity are steep functions of $r/\rs$, and hence more sensitive to the precise value of $\cv$.
The effect on these quantities is also stronger at large overdensities, but unlike the $M-T$ relation, the systematic variation of concentration would yield a noticeable imprint on the $F-T$ and $L-T$ relations at $\Delta=200$.

On the other hand, the effective polytropic index controls the slope of the gas density profile.
According to~(\ref{eqCG}), larger masses, which imply lower concentrations, also mean lower effective polytropic index.
The gas density profile becomes increasingly steep, which boosts the central baryon fraction and the X-ray luminosity, increasing the emission-weighted temperature only slightly and leaving the the total mass mostly unaffected.
Given the small variation of the polytropic index throughout the interesting mass range, the effect is never larger than a factor of two, comparable to or smaller than that of the concentration, but it always acts in the opposite sense.
}

Finally, we would also like to note that, apart from the normalization, our model can estimate the scatter around the average scaling relations by assuming that it arises completely from the scatter in the mass-concentration relation.
Such estimate, indicated by the dotted lines in Figure~\ref{figUncorr}, has been obtained by combining the value $c'=1.3\cv$ with the effective polytropic index $\gamma'=\gamma(\cv/1.3)$, and $c''=\cv/1.3$ with $\gamma''=\gamma(1.3\cv)$, where in both cases $\gamma(c)$ has been computed according to equation~(\ref{eqCG}).
Comparing to Figure~\ref{figCorr}, it seems that the internal structure of clusters (accounted for by the factors $\Ymt$, $F_\Delta$ and $\Ylt$) is indeed responsible for a significant fraction of the scatter in the observed scaling relations, as recently suggested by \citet{OHara_05}.
Note, however, that in our case the differences in internal structure are obviously not related to the presence of a cool core or the action of any external source of energy, but rather to the different formation histories of each object.

%___________________________________________________

\subsection{Comparison with previous work}

%__________________________________
\begin{table}
\caption{
Normalization $M_0^\Delta$ (in $10^{14}~h^{-1}$~\Msun) and logarithmic slope $\alpha$ of the $M-T$ relation reported in previous numerical studies.
}
\begin{center}
\begin{tabular}{lrcl}
\hline {\sc Reference} & $\Delta$~ & $M_0^\Delta$ & ~$\alpha$ \\ \hline
\citet{NFW95}       &  200 & 6.14 & 1.5  \\
\citet{EMN96}       & 2500 & 7.85 & 1.5  \\
                    &  500 & 7.84 & 1.5  \\
\citet{BN98}        &  200 & 8.68 & 1.5  \\
\citet{Pen98}       &  200 & 7.28 & 1.5  \\
\citet{Eke98}       &  100 & 6.12 & 1.5  \\
\citet{Yoshikawa00} &  100 & 5.56 & 1.5  \\
\citet{ME01}        &  500 & 8.18 & 1.52 \\ % Be careful with h=.8 typo in Table 1
                    &  200 & 7.65 & 1.54 \\
\citet{Muanwong02}  &  200 & 13.9 & 1.5  \\
\hline
\end{tabular}
\end{center}
\label{tabMTsims}
\end{table}
%__________________________________

The mass-temperature relation in the absence of radiative processes has been extensively studied by means of cosmological numerical simulations.
A summary of previous results is given in Table~\ref{tabMTsims}.
Slopes consistent with $3/2$ are found in most studies, with normalizations showing a relatively low scatter around $M_0^\Delta\sim(7-8)\times10^{14}~h^{-1}$~\Msun.
Similar results are obtained when cooling and stellar feedback are considered \citep[e.g.][]{Borgani04}, although there is a trend towards steeper slopes and lower normalizations, in better agreement with observational data.

The $M-T$ relation predicted by our model (see Table~\ref{tabFiducial}) is also considerably lower than the values found in previous experiments based on purely adiabatic gasdynamics.
We think that this is to a great extent a resolution effect, coupled to the use of an entropy-conserving scheme to solve the SPH equations.

Most if not all of the earlier work on the adiabatic scaling relations of galaxy clusters relied on the traditional formulation of SPH \citep{Lucy77,GingoldMonaghan77}.
It has been recently shown \citep[e.g.][]{Gadget02,Ascasibar03,OShea05} that poor entropy conservation leads to spurious entropy losses in the cluster cores, and thus previous codes tend to systematically overestimate the central density and underestimate the central gas temperature.

On the other hand, lack of resolution results in artificially flattened density and temperature profiles, i.e. nearly isothermal and isentropic cores.
Actually, some studies \citep[e.g.][]{ME01,Muanwong02} even report decreasing temperature profiles towards the centre.
This is in strong disagreement with our results, as well as with those of independent numerical work based on high-resolution Eulerian simulations \citep[e.g.][]{Loken02}, in which the temperature profile in the absence of radiative processes is also found to decrease monotonically with radius.
Under such conditions, the emission-weighted temperature (which is biased towards the central, dense and X-ray bright regions of the cluster) is larger than the mass-weighted average, and thus the resulting normalization of $M-T$ relation becomes considerably lower when $\Tx$ is used \emph{and} the central parts of the objects under study are well resolved.

Concerning the $\Lx-\Tx$ relation, there is relatively little numerical work based on adiabatic gasdynamical simulations.
This is also a reflection of the stringent resolution requirements, due to the fact that a significant fraction of the X-ray photons are expected to be produced in the innermost regions.
A further problem affecting cosmological numerical experiments is that the smallest objects are typically resolved with less particles, so their bolometric X-ray luminosity is underestimated and the resulting $L-T$ relation is artificially steepened \citep[see e.g.][]{BN98,Yoshikawa00,Yepes_04}.
Most of our objects (see Table~\ref{tabSims}) have more than $10^5$ gas particles, and in principle they should not be severely affected by this problem \citep[see e.g.][]{Borgani02,Borgani_05}.
Nevertheless, it is always wise to bear this consideration in mind when drawing conclusions from numerical data.

%__________________________________
\begin{table}
\caption{$L-T$ relation found in previous simulations, with $L_0^\Delta$ expressed in $10^{43}~h$~erg~s$^{-1}$.}
\begin{center}
\begin{tabular}{lrcl}
\hline {\sc Reference} & $\Delta$~ & $L_0^\Delta$ & ~$\alpha$ \\ \hline
\citet{NFW95}       &  200 & 4.61 & 2 \\
\citet{BN98}        &  200 & 2.67 & 2 \\
\citet{Eke98}       &  100 & 0.85 & 2 \\
\citet{Bialek01}    &  500 & 2.44 & 2.02 \\
\hline
\end{tabular}
\end{center}
\label{tabLTsims}
\end{table}
%__________________________________

Table~\ref{tabLTsims} shows several fits to the adiabatic $L-T$ relation reported in the literature.
The normalizations are in this case broadly consistent with our results (Table~\ref{tabFiducial}), although the scatter between different estimates is extremely large.
This is not entirely unexpected, given the sensitivity of the X-ray luminosity to the details of the gas density profile in the central regions.

Finally, the baryonic content of galaxy clusters has been recently investigated by \citet{Kravtsov05}.
For their adiabatic simulations, they find $F_{2500}=0.85\pm0.08$ and $F_{500}=0.94\pm0.03$ in units of the cosmic value.
As noted by these authors, the baryon fractions obtained with the Eulerian code {\sc ART} \citep{ARThydro02} are systematically less concentrated than those obtained with {\sc Gadget}, even when entropy conservation is enforced, but nevertheless the cumulative baryon fractions beyond $r_{2500}$ are about about $3-5$ per cent higher.
Since the results of \citet{Kravtsov05} are based on a subset of the cluster sample studied here, the interested reader is referred to that paper for an extensive comparison between both codes.

Our results are also compatible with the baryon fraction measured by \citet{Ettori06} in their non-radiative runs, when the standard implementation of the artificial SPH viscosity is used.
As pointed out by these authors, it is interesting how an improved scheme may lower the baryon fraction in the innermost regions by about 15 per cent.
Although not dramatic, it seems clear that the details of the numerical technique do have a measurable impact on the predicted radial profiles near the centre, as well as on the scaling relations at high overdensities.
These relatively small discrepancies between different algorithms should nevertheless be considered as part of the theoretical uncertainty.

%--------------------------------------------------------------------------
  \section{Observations}
  \label{secObs}
%--------------------------------------------------------------------------

%__________________________________
\begin{figure*}
  \centering \includegraphics[width=15cm]{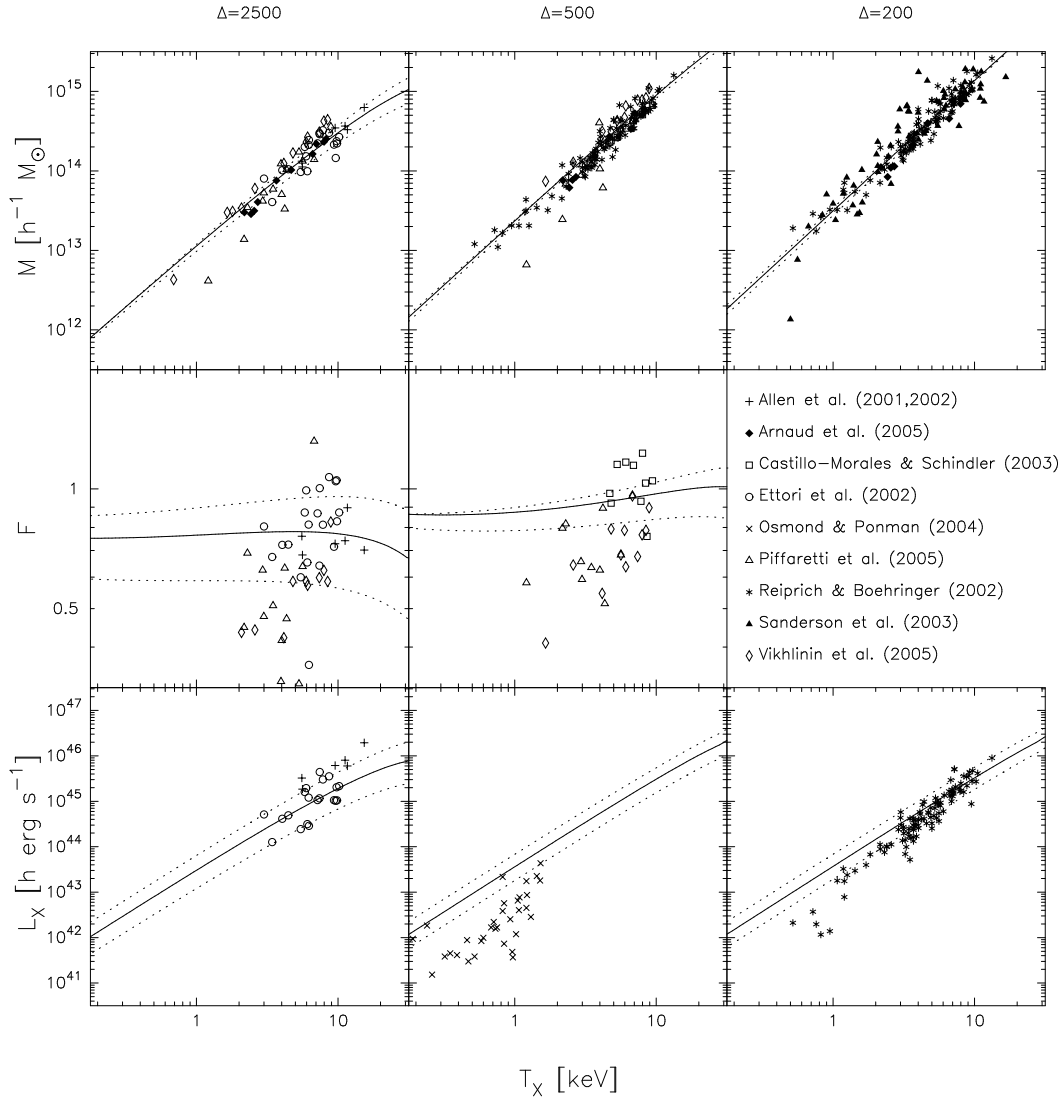}
  \caption{
Observed $M-\Tx$, $F-\Tx$ and $\Lx-\Tx$ scaling relations, compared to our theoretical prediction (solid lines) and the one-sigma scatter expected from $\Delta\cv/\cv\simeq0.3$ (dotted lines).
}
  \label{figObs}
\end{figure*}
%__________________________________

It is the aim of the present work to provide a sound theoretical prediction of the cluster scaling relations when only gravity, adiabatic gasdynamics and shock-wave heating act on the intracluster medium.
In real life, additional physical processes, such as radiative cooling, energy injection by stars and AGN or thermal conduction, may play an important role or even determine the exact form of the scaling relations.
However, the influence of all these phenomena outside the central regions is expected to be relatively small when compared to shock heating, especially for the most massive systems.

Therefore, one may expect a priori that the adiabatic scaling relations roughly match the observed ones.
Departures would measure the effect of additional physics, and in principle should be more noticeable at high overdensities and for low-temperature systems.

Our predictions for the adiabatic case are compared with observational data in Figure~\ref{figObs}.
We find a fairly good agreement, both in shape and, to some extent, scatter, with the observed $M-T$ relation at different overdensities.
Only the dataset from \citet{Piffaretti05} is not well described by our model.
From visual inspection of Figure~\ref{figObs}, though, it seems that the dark matter masses inferred for some of these systems are lower not only than our results but also than the other observations.

\Referee{
On the other hand, the luminosity-temperature relation observed for galaxy clusters is roughly consistent with our theoretical prediction, but the slope is appreciably steeper.
As one approaches the group regime, real systems can be one order of magnitude less bright than the model.
}
The X-ray luminosity is much more sensitive to the details of the central parts than the total mass or the emission-weighted temperature, and thus it is not surprising that the $L-T$ relation deviates from the adiabatic prediction more significantly than the mass-temperature relation.

The lower X-ray emissivity seems to be intimately connected to the shape of the gas density profile.
Our models correctly predict that the baryon fraction should be an increasing function of radius, and such a trend is clearly consistent with the observational data.
However, they also predict that, at a given overdensity, the baryon fraction should be roughly independent on cluster mass.

This is blatantly at odds with observations.
As noted by \citet{Vikhlinin_05scal}, the conversion of gas into stars is a crucial factor that should be taken into account.
Actually, it would be very interesting to measure whether it could completely explain the observed central baryon depletion on its own, or by the contrary some other physical mechanism (e.g. heating) must be invoked in order to explain the observed density and temperature profiles \citep[see e.g.][for a recent discussion on this issue]{Borgani_05}.

In any case, radiative processes must obviously affect the physical properties of real clusters to some extent.
As recently shown by \citet{OHara_05}, the scaling relations of clusters with and without a cool core are clearly offset from each other, and the observed scatter can be significantly reduced by introducing the peak strength (characterized by the central X-ray surface brightness) as an additional parameter.

Although such a parameter would measure a mixture between the intensity of cooling and the internal structure of the halo, a visual comparison between the simulation results plotted in Figure~\ref{figUncorr} and the observational data shown in Figure~\ref{figObs} suggests that observed systems display a somewhat larger scatter than our simulated clusters, which points in the direction that both processes may have a comparable contribution to the total scatter.
\Referee{
Measurement errors, most notably for cluster mass estimates, also contribute to the scatter in the observed scaling relations, although they have been reported to be relatively small compared to the intrinsic scatter \citep{OHara_05}. We have also neglected redshift evolution, which can modify the observed masses and luminosities by a factor $H(z)/H_0$, which for a $\Lambda$CDM universe amounts to about 40 per cent at $z=0.2$.
}
A rigorous statistical analysis (and a larger dataset) would be required in order to make a quantitative assessment.

%--------------------------------------------------------------------------
  \section{Conclusions}
  \label{secConclus}
%--------------------------------------------------------------------------

In this paper, the scaling relations between gas and dark matter mass, X-ray luminosity and emission-weighted temperature of galaxy groups and clusters have been investigated from a theoretical point of view.
As a starting point, we have considered a relatively simple case in which the influence of radiative processes on the observable properties of the ICM gas has been completely neglected.
Our estimates of the adiabatic scaling relations have been computed from the polytropic models described in \citet{Ascasibar03}, based on the results of high-resolution gasdynamical simulations.

Our main conclusions can be summarized as follows:

\begin{enumerate}
\item Dark matter haloes are well known not to scale self-similarly, but according to a certain mass-concentration relation.
We find that the effective polytropic index of the gas also varies systematically with mass, and propose the phenomenological fit $\gamma=1.145+0.005\,c_{200}$, equation (\ref{eqCG}), to model the dependence of $\gamma$ on the concentration $c$ of the dark matter halo.
\item Given $c(M)$ and $\gamma(c)$, the whole structure of the ICM is fully specified by our model.
It turns out that the effects of the varying polytropic index and concentration tend to cancel out at all overdensities, yielding scaling relations that are well described by simple power laws whose exponents coincide with the self-similar prediction, and whose normalizations are well fitted by a `typical' $\cv\simeq8$.
\item Our model provides an excellent match to numerical data.
The normalization of the M-T relation is significantly lower than previous values reported in the literature, which we attribute to a resolution effect.
The scaling of the baryon fraction and the $L-T$ relation are broadly consistent with independent numerical work.
\item Additional physics (most notably, radiative cooling and star formation) has an important effect on the density and temperature profiles of real clusters.
The $M-T$ relation is not severely affected, but the baryon fraction observed in low-mass systems is considerably below our theoretical prediction.
This results in a lower X-ray luminosity, and it is ultimately responsible for the steepness of the observed $L-T$ relation.
On the other hand, the precise strength of cool cores (which cannot form in our simulations) seems to increase the scatter around the average scaling relations.
\end{enumerate}

%---------------------------------------------------------------

\section*{Acknowledgments}

This work has been partially supported by the \emph{Plan Nacional de Astronom\'\i a y Astrof\'\i sica} (AYA2003-0973), the \emph{Acciones Integradas Hispano-Alemanas} (HA2000-0026), the \emph{Deutscher Akademischer Austausch Dienst} and NASA grants G02-3164X and G04-5152X.
We thank the CIEMAT, the \emph{Forschungszentrum J\"ulich} and the \emph{Astrophysikalisches Institut Potsdam} for kindly allowing us to use their supercomputer facilities to carry out the numerical simulations used in this article.

%--------------------------------------------------------------------------

%Observations \citep{ASF01,ASF02omega,Arnaud05,CastilloSchindler03,EttoriGrandiMolendi02,OsmondPonman04,Piffaretti05,ReiprichBoehringer02,Sanderson03,Vikhlinin_05scal}

\bibliographystyle{mn2e}
\bibliography{bibtex/DATABASE,bibtex/PREPRINTS}
%\bibliography{bibtex/DATABASE,bibtex/PREPRINTS,bibtex/raul}

\end{document}